\documentstyle[11pt,moriond]{article}
\input epsf

\def\be{\begin{equation}}
\def\ee{\end{equation}}
\def\bea{\begin{eqnarray}}
\def\eea{\end{eqnarray}}

\def\ltsima{$\; \buildrel < \over \sim \;$}
\def\lsim{\lower.5ex\hbox{\ltsima}}
\def\gtsima{$\; \buildrel > \over \sim \;$}
\def\gsim{\lower.5ex\hbox{\gtsima}}
\def\bi#1{\hbox{\boldmath{$#1$}}}
\begin{document}
\vspace*{4cm}
\title{Polarization of Microwave Background: Statistical 
and Physical Properties}

\author{Uro\v s Seljak $^1$, Matias Zaldarriaga $^2$}

\address{$^{1}$
Harvard Smithsonian Center For Astrophysics, Cambridge, MA 02138, USA\\
$^{2}$
MIT, Cambridge, MA 02139, USA}

\maketitle\abstracts{We discuss statistical and physical properties 
of cosmic microwave background polarization, both in Fourier and 
in real space. The latter allows for a
more intuitive understanding of some of the geometric signatures. 
We present expressions that relate electric and magnetic type of 
polarization to the measured Stokes parameters in real space and 
discuss how can be constructed locally. We discuss necessary 
conditions to create polarization and present maps and correlation 
functions for some typical models.}

\section{Introduction}
Cosmic microwave background (CMB) anisotropies offer one of the best probes
of early universe, which could potentially lead to a precise
determination of a large number of cosmological parameters 
\cite{jungman,1.bet,2.zss}.
The main advantage of CMB versus more local probes of large-scale
structure is that the fluctuations were created at an epoch when the
universe was still in a linear regime. While this fact has long been
emphasized for temperature anisotropies ($T$), the same holds also
for polarization in CMB and as such it offers the same advantages
as the temperature anisotropies in the determination of cosmological
parameters. The main limitation of polarization is that it is
predicted to be
small: theoretical calculations show that CMB will be polarized
at 5-10\% level on small angular scales and much less than that
on large angular scales. Future CMB missions (MAP, Planck) 
will have sufficient sensitivity that even such low signals
will be measurable. This will allow one to exploit the wealth 
of information present in the polarization. 

Recent work has emphasized the rich geometrical structure present 
in polarization \cite{2.uros,2.kks,2.spinlong}. 
In particular, linear polarization has been 
decomposed into electric ($E$) and magnetic ($B$) types,
which transform as scalars and pseudoscalars, respectively. 
With polarization there are three additional 
power spectra that can be measured, in addition to $E$ and $B$ 
autocorrelation there is also $E$ and $T$ cross-correlation. 
Each of these can provide unique
information about our universe. Most of this work has developed the
formalism by using multipole expansion on a sphere or on a plane. 
Here we will 
develop some of the properties of polarization fields $E$ and $B$ 
directly in real space, which 
allows for a more intuitive understanding of their geometrical properties.

\section{Statistical description of polarization}
The CMB radiation field is characterized by a $2\times 2$ intensity
tensor $I_{ij}$. The intensity tensor is a
function of direction on the sky $\bi n$ and  two directions
perpendicular to $\bi n$ that are  used to define its components
(${\bf \hat e}_1$,${\bf \hat e}_2$).
The Stokes parameters $Q$ and $U$ are defined as
$Q=(I_{11}-I_{22})/4$ and $U=I_{12}/2$, while the temperature
anisotropy is
given by $T=(I_{11}+I_{22})/4$ (the factor of $4$ relates fluctuations
in the intensity to those in the temperature).
In color plates
we represent the
polarization using ``vectors'' with magnitude $P=\sqrt{Q^2+U^2}$ that form
an angle $\alpha={1\over 2}\arctan({U \over Q})$ with ${\bf \hat e}_1$.
These are not real vectors 
since they map into themselves after $180^{\circ}$ rotation, which 
is why we do not assign them a direction.
In principle the fourth
Stokes parameter $V$ that describes circular polarization would also
be needed, but in cosmology it can be ignored because it cannot
be generated through Thomson scattering. 
While the temperature is invariant
under a right handed rotation in the plane perpendicular to direction
$\bi n$,
$Q$ and $U$ transform under rotation by an angle $\psi$ as
\begin{eqnarray}
Q^{\prime}&=&Q\cos 2\psi  + U\sin 2\psi  \nonumber \\
U^{\prime}&=&-Q\sin 2\psi  + U\cos 2\psi
\label{QUtrans}
\end{eqnarray}
where ${\bf \hat e}_1^{\prime}=\cos \psi\ {\bf \hat e}_1+\sin\psi\
{\bf \hat e}_2$
and ${\bf \hat e}_2^{\prime}=-\sin \psi\ {\bf \hat e}_1+\cos\psi\
{\bf \hat e}_2$.
The Stokes parameters are not invariant under
rotations in the plane
perpendicular to $\bi n$. For this
reason it is more convenient to work with scalar and pseudoscalar
polarization fields 
$E(\bi n)$ and $B(\bi n)$, 
respectively, which are invariant under rotations.
In the small scale limit $\bi n$ is close to $\hat{\bi{z}} $
and we can parametrize the direction in the sky with two-dimensional angle 
$\bi{\theta}$ relative to a fixed coordinate system perpendicular 
to $\hat{\bi{z}} $. The two rotationally invariant fields can be written in terms of 
the Stokes parameters as
\begin{eqnarray}
E(\bi{l})&=&\int d^2{\bi{\theta}}\
[Q(\bi{\theta})\cos(2\phi_l)+U(\bi{\theta})\sin(2\phi_l)] \ e^{-i{\bi l}
\cdot {\bi \theta}} \nonumber \\
B(\bi{l})&=&\int d^2{\bi{\theta}}\
[U(\bi{\theta})\cos(2\phi_l)-Q(\bi{\theta})\sin(2\phi_l)] \ e^{-i{\bi l}
\cdot {\bi \theta}},
\end{eqnarray}
Here $E(\bi{l})$ and $B(\bi{l})$ are components of the two scalar 
fields in Fourier space. To obtain them in real space we can perform 
a Fourier transform
\begin{eqnarray}
E(\bi{\theta})&=&(2\pi)^{-2}\int d^2{\bi l}\
e^{i{\bi l}{\bi \theta}}\ E({\bi l}) \nonumber \\
B(\bi{\theta})&=&(2\pi)^{-2}\int d^2{\bi l}\
e^{i{\bi l}{\bi \theta}}\ B({\bi l}).
\label{ebreal1}
\end{eqnarray}
These two quantities  describe completely the polarization field.
They can be expressed directly in terms of real space quantities 
$Q(\bi{\theta})$ and $U(\bi{\theta})$
as 
\begin{eqnarray}
E(\bi{\theta})&=&-\int d^2{\bi \theta}^{\prime}\
\omega(\tilde \theta)\ [Q({\bi \theta}^{\prime})
\cos(2\tilde\phi) - U({\bi \theta}^{\prime})
\sin(2\tilde\phi)]\nonumber \\
&=&-\int d^2{\bi \theta}^{\prime}\
\omega(\tilde \theta)\ Q_r({\bi \theta}^{\prime}) \nonumber \\
B(\bi{\theta})&=&-\int d^2{\bi \theta}^{\prime}\
\omega(\tilde \theta)\ [U({\bi \theta}^{\prime})
\cos(2\tilde\phi) + Q({\bi \theta}^{\prime})
\sin(2\tilde\phi)]\nonumber \\
&=&-\int d^2{\bi \theta}^{\prime}\
\omega(\tilde \theta)\ U_r({\bi \theta}^{\prime}).
\label{ebreal2}
\end{eqnarray}
The variables $(\tilde \theta,\tilde \phi)$ are the polar coordinates
of the vector $\bi{\theta}-\bi{\theta}^{\prime}$.
In equation (\ref{ebreal2})
$Q_r$ and $U_r$ are the Stokes parameters in the
polar coordinate system centered at $\bi{\theta}$. For example,
if $\bi{\theta}$ is zero, $Q_r=\cos 2\phi^{\prime}\
Q(\bi{\theta}^{\prime}) - \sin 2\phi^{\prime}\
U(\bi{\theta}^{\prime})$ and $U_r=\cos 2\phi^{\prime}\
U(\bi{\theta}^{\prime}) + \sin 2\phi^{\prime}\
Q(\bi{\theta}^{\prime})$. The window can be shown to be
 $\omega(\theta)=1/\pi \theta^2\;
(\theta\neq 0)$, $\omega(\theta)=0 \;
(\theta= 0)$. 

From equation \ref{ebreal2} we can understand
how to obtain two rotationally invariant quantities out
of $Q$ and $U$ directly in real space: 
to obtain $E({\bi \theta})$ and $B({\bi \theta})$
we radially integrate over circles centered at ${\bi
\theta}$ the values of 
$Q_r$ and $U_r$, respectively.
Each circle is weighted by $\omega({\tilde \theta})$. By
construction these two quantities are rotationally invariant: the
Stokes parameters $Q_r$ and $U_r$ do not depend on the coordinate
system, as they are defined relative to the
$\bi{\theta}-\bi{\theta}^{\prime}$
vector. The weight function $\omega$ is also rotationally invariant.
We are
giving the same weight to all the points in each circle and we are
using the Stokes parameters defined in their {\it natural} coordinate
system, defined parallel and
perpendicular to the line joining the two points (denoted with $r$ in
equation \ref{ebreal2}). 
The variable $B$ is clearly a pseudoscalar because it is the average
of $U_r$ and $U_r$ changes sign under parity. Note that the window 
extends out to infinity and so the quantities $E$ and $B$ are nonlocal.
This is not the only possible choice: one can construct finite extent
versions of $E$ and $B$ from $Q$ and $U$ using a compensated window
such that $\int \theta d\theta\ g(\theta)=0$, e.g. 
$\tilde{E}=\int E g(\theta)d^2 \theta$ and similarly for $B$. The 
relation between these averaged $\tilde{E}$ and $\tilde{B}$
and $Q$, $U$ is still given by equation \ref{ebreal2} using
\begin{equation}
\omega(\theta)= -g(\theta)+{ 2 \over \theta^2 } \int_0^{\theta}
d\theta ' \theta ' g(\theta ').
\end{equation}
We can therefore construct pure $E$ or $B$ quantities which are 
obtained from an integral over a finite region and test the 
geometrical structure of polarization without measuring the 
whole sky, as long as the measured field is contiguous. Note 
that for these real space quantities there is no need to worry 
about mode coupling because of incomplete sky coverage, which 
would for example mix some $E$ into $B$ in Fourier space. 

\begin{figure}
\begin{center}
\leavevmode
\epsfxsize=4in \epsfbox{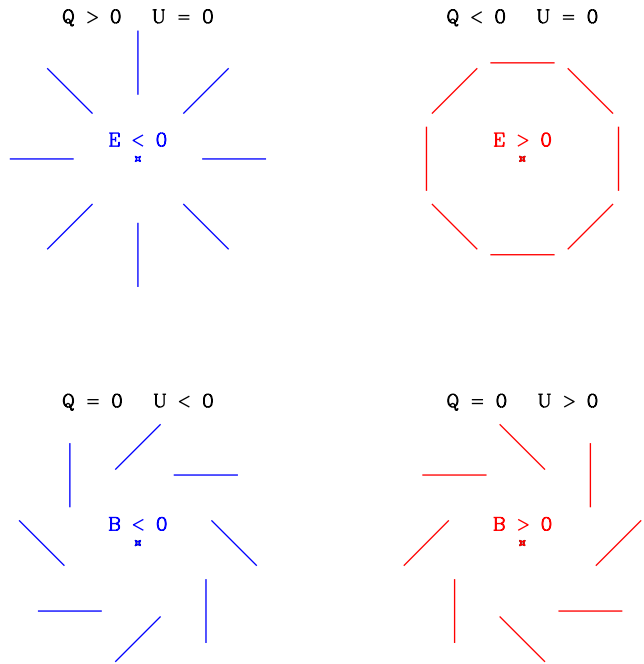}
\end{center}
\caption{Polarization patterns that lead to positive and negative
values of the $E$ and $B$ fields. The Stokes parameters are measured
in the polar coordinate system centered at the cross. All four
patterns are invariant under rotation but the two patterns that
generate $B$ are not invariant under reflections.}
\label{2.ebpatt}
\end{figure}

Let us further explore properties of $E$ and $B$ type polarizations
in real space. For simplicity we will continue to use 
small scale limit.
Plate 1 shows the $E$ polarization field and the
polarization vectors for a typical (standard CDM) model.
There is only $E$ type polarization associated with this model, 
because
this is the only pattern that is produced by density perturbations and
we did not include a stochastic background of gravity waves. We discuss 
this point further below.
In figure 
\ref{2.ebpatt} and plate 1 we can see that hot spots of
the $E$ map correspond to points with tangential polarization pattern
(negative $Q_r$). Radial polarization pattern is found around the 
cold spots of $E$. 
From equation \ref{ebreal2} we 
can read directly the
relations between the pattern of polarization and the sign of $E$ and
$B$. Note that there is a formal similarity between weak lensing and 
polarization: $E$ plays a role of convergence $\kappa$, while $Q$ and
$U$ can be viewed as the two shear components $\gamma_1$ and $\gamma_2$.
Since weak lensing is only important for scalar perturbations $B$ mode is 
not excited, which provides a consistency check. The sign 
convention for $E$ 
agrees with weak lensing \cite{Kaiser92}: positive mass density (positive 
$\kappa$) generates a tangential pattern of shear, so likewise 
positive $E$ gives rise to a tangential polarization pattern.

To obtain $B$ type polarization we can rotate all polarization ``vectors''
by $45^{\circ}$. {\it Hot} and {\it cold} spots of the
$B$ field correspond to places where the polarization vectors
circulate around in opposite directions (figure \ref{2.ebpatt} and 
plate 2). 
It is clear from this figure
that such polarization pattern is not invariant under reflections 
(parity transformation). This is the main 
distinction between $E$ and $B$ type of polarization:
under parity operation $E$ transforms 
as a scalar
and $B$ as a pseudoscalar.

\begin{figure}[t]
\begin{center}
\leavevmode
\epsfxsize=4in \epsfbox{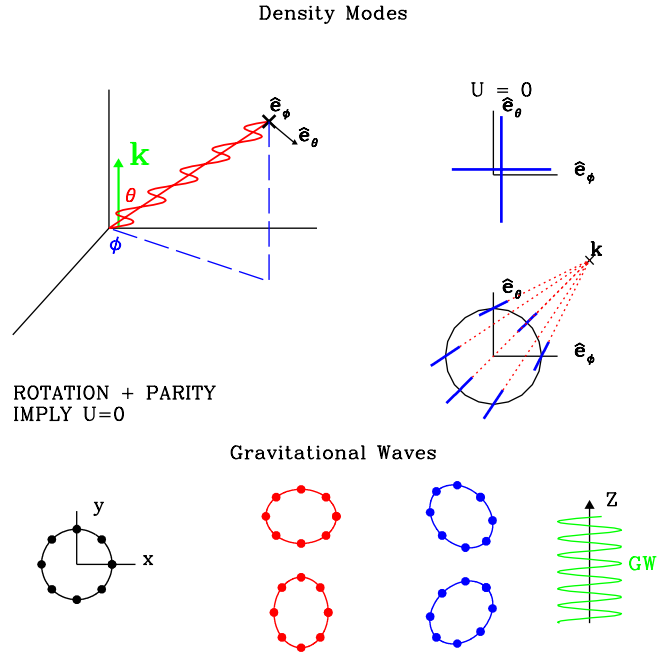}
\end{center}
\caption{A single mode of density perturbations has symmetries under
rotation around the $\hat{\bf k}$ axis and reflection around any plane
containing it. Only $Q$ Stokes parameter can be present in this
reference frame, implying 
no $B$ polarization is created (see the text). Gravitational
waves do not have these symmetries as illustrated by the deformation
suffered by a ring of test particles as wave propagating in direction
$\hat{\bf k}$ passes by. There is a particular direction associated 
with the direction of the stretching and there is no 
rotational or parity symmetry. The same is true for vector modes, which
induce velocities perpendicular to the wave direction.}
\label{2.densp2}
\end{figure}

We can now address the polarization pattern induced by different
types of perturbations. In particular, we want to show 
why scalar perturbations 
cannot induce $B$ type of polarization. Without loss of generality 
we may consider only one 
Fourier mode of scalar perturbations at a time, since 
an arbitrary field can always be expanded in a linear superposition of
Fourier modes. Let us assume we are observing in direction $\bi n$, while 
the mode is oriented in direction $\hat{\bf k}$. 
In the plane perpendicular 
to $\bi n$ we can define a local coordinate system with axis aligned 
parallel and perpendicular to the plane that contains both 
$\hat{\bf k}$ and $\bi n$. Scalar modes have
a rotation symmetry around $\hat{\bf k}$, which implies that $U=0$ in this frame 
(figure \ref{2.densp2}).
Let us now look at the polarization pattern in the circle 
around direction $\bi n$ 
in the plane perpendicular to it. Polarization is directed either
in the direction towards $\hat{\bf k}$ or perpendicular to this direction,
since only $Q$ is generated in the frame determined by that direction 
(for simplicity only radial case is depicted in figure \ref{2.densp2}).  
Polarization is invariant under the 
reflection across the axis determined by the 
$\hat{\bf k}$ and $\bi n$ line, 
because polarization amplitude only depends on the angle between 
$\hat{\bf k}$ and $\bi n$. Therefore any integration around this circle will
produce only $E$ and not $B$. Since both $E$ and $B$ are obtained by 
radial averaging of contributions from individual circles and $B$ 
vanishes for each for each of them it follows that $B=0$ in general. 
By construction this is coordinate independent, 
hence one cannot produce $B$ polarization with 
scalar perturbations in general. 

For tensor perturbations this argument no longer 
applies, as shown in figure \ref{2.densp2}. Here the amplitude 
of perturbation depends also on the orientation of the stretching and
squeezing of space in the direction perpendicular to the direction of 
plane propagation (figure \ref{2.densp2}). 
There is no parity symmetry and both $E$ and $B$
will be generated in general. This is also true for vector modes. For 
these modes velocity is perpendicular to the
direction of wave propagation. Its direction in the plane perpendicular 
to $\hat{\bf k}$ breaks the symmetry and generates $B$ modes.

\section{Generation of Polarization}
The standard picture of big bang model is that the universe was hot 
in the beginning and was expanding and cooling thereafter. 
At high temperatures electrons and protons were free in the 
universe, because photons were sufficiently energetic to prevent
any recombination to survive. 
Once the energy of most of the photons dropped below
13.6eV ionization was no longer possible and at 
that time (at energies around 0.3eV) the universe recombined and became 
transparent for photons. The recombination was very rapid. Before 
recombination the density of free electrons was very high, so that 
the photon mean free path was very short. Together with electrons 
they formed a single fluid, with pressure provided by the
photons and inertia by the baryons. This fluid supported the
analog of acoustic oscillations where both
the density  and velocity were oscillating
functions of time \cite{2.uroscmb,2.wayneth}. 
The density  was proportional to  $\cos (c_s
k\tau+\delta)$, while the velocity to $\sin(c_s k\tau+\delta)$, where
$c_s$ is the sound speed and the phase $\delta$ depends on initial 
conditions and is in general time dependent.

To generate polarization we need 
Thomson scattering between photons and electrons. 
Electric field incident on the electron causes 
this electron to oscillate in the direction of the field and 
electron radiates according to the dipole emission
formula. Looking for example perpendicular to the incoming 
light direction only one
polarization of incident light will be seen to cause electron to oscillate, 
the one perpendicular to the plane containing the incoming and outgoing
directions. Dipole 
radiation emits preferentially perpendicular to
the direction of oscillation, 
so the light we observe will be polarized. To get the total effect we
need to integrate over the light intensity in all the directions.
If the radiation incident on the
electron is isotropic there will be no net polarization generated
after the scattering, simply by symmetry. Dipole distribution also
does not generate polarization.
For example, consider the case where the
intensity of the incident radiation is higher from the top and lower from th
e
bottom, with the average intensity incident from the sides.
Observing from above the plane
scattered light from photons incident from either the
bottom or top of the plane 
will be polarized in the horizontal direction, while
that coming from the sides will be polarized in the vertical direction.
Adding up all the components we find that horizontal and vertical 
components are equal, implying no polarization. We therefore need 
a quadrupolar pattern of
intensity, in which case 
there is an 
excess of horizontal polarization from both top and bottom, which 
does not cancel the smaller vertical polarization from the sides.

Since we need Thomson scattering it is clear that polarization 
cannot be generated after recombination of electrons and protons 
into hydrogen, when the universe becomes transparent for the photons.
We still need to show however that there is sufficient quadrupole moment 
prior to last scattering, because only then will polarization be 
generated. In the electron rest frame one sees photons coming 
from distances of the order of a mean free path $\lambda_p$. The photons that are
scattered off a given electron come
from places where the fluid has velocity ${\bf {v}
}$ and because of the tight coupling between photons and
electrons the photon distribution function
has a dipole term
$T_1=\hat {{\bf n}}\cdot {\bf {v}}$.
To generate quadrupole moment one requires a gradient in the 
velocity field across the mean free path, 
$T_2=\lambda _pn^in^j\partial _iv_{j}$ in the rest frame of the electron.
But since the mean free path was so short prior to recombination the 
induced quadrupole was very small, until the last moment during recombination
when mean free path started to grow. This implies that 
the mean free path is of the order of the duration 
of recombination and a detailed analysis gives \cite{zalhar} 
\begin{equation}
(Q+iU)(\hat {{\bf n}})\approx 0.17\Delta \tau _{*}{\bf m}^i{\bf m}^j\partial
_iv_j|_{\tau _{*}},
\label{polapprox}
\end{equation}
where $\Delta \tau _{*}$ is the width of the last scattering surface,
happening at $\tau _{*}$.
The appearance of ${\bf m}=\hat {{\bf e}}_1+i\hat {%
{\bf e}}_2$ in equation \ref{polapprox} assures
that $(Q+iU)$ transforms correctly under rotations of $(\hat {{\bf e}%
}_1,\hat {{\bf e}}_2)$ (equation \ref{QUtrans}).

\begin{figure*}
\begin{center}
\leavevmode
\epsfxsize=4in \epsfbox{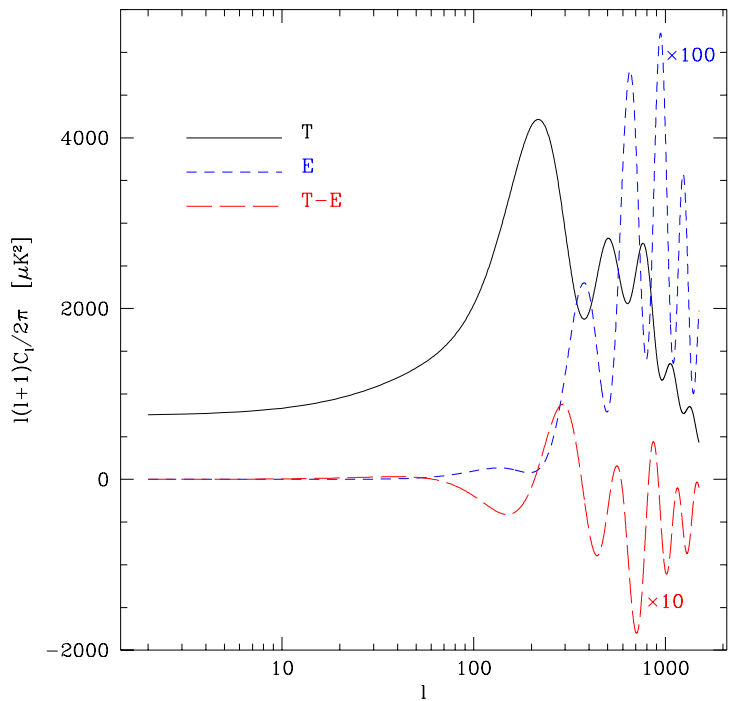}
\end{center}
\caption{ Power spectra for COBE normalized SCDM. The $E$ and the
$T-E$ spectra have been rescaled for convenience.}
\label{figclcdm}
\end{figure*}

Figure \ref{figclcdm} shows the temperature, $E$ polarization and
$T-E$ cross-correlation power spectra. 
The oscillations in the spectra can be understood
with the tight coupling approximation described above. 
At the time of recombination waves of different wavevectors $k$ 
are at different phases in their oscillations. The phase 
only depends on the amplitude of the wavevector, so all the modes
with the same amplitude of wavevector have the same phase. For
particular values of this amplitude all the corresponding modes 
vanish. A given physical scale is related to the corresponding 
angular scale via the angular diameter distance, but the 
correspondence is not perfect because of 3-d to 2-d projection. 
If it were perfect the power spectrum of polarization would vanish
at some values, but in reality it does not (figure \ref{figclcdm}). 
Nevertheless, the oscillations in polarization are much more 
pronounced than in temperature, because the latter receives 
contributions both from density and velocity and the two are
out of phase with each other, so that even if one vanishes for some 
value of $k$ the other does not.

On large scales 
polarization is strongly suppressed. Correlations over large angles can
only be created by the long
wavelength perturbations, but these cannot produce a large
polarization because of the tight coupling between photons and
electrons prior to recombination: only wavelengths that are small enough to
produce anisotropies over the mean free path of the photons will give
rise to a significant quadrupole in the temperature distribution, and
thus to polarization.
On small scales ($l \geq  1000$) both polarization and temperature 
anisotropies are suppressed
by photon diffusion. Here the wavelength of the perturbation becomes
comparable to the mean free path of the photons prior to
recombination. Photons can diffuse out of density peaks smaller
than a diffusion length, thereby erasing the anisotropies.
Note also that there is an extra gradient in polarization relative
to temperature (equation \ref{polapprox}), which explains 
why polarization amplitude peaks at a smaller
angular scale. The fact that the polarization field has relatively
more small scale power is particularly 
evident when we compare the the $T$ and $E$
fields in 
plates 1 and 3.
Finally, temperature polarization cross-correlation oscillates just 
like the other two spectra, but can be either positive or negative 
(figure \ref{figclcdm}). 

\begin{figure*}
\begin{center}
\leavevmode
\epsfxsize=4in \epsfbox{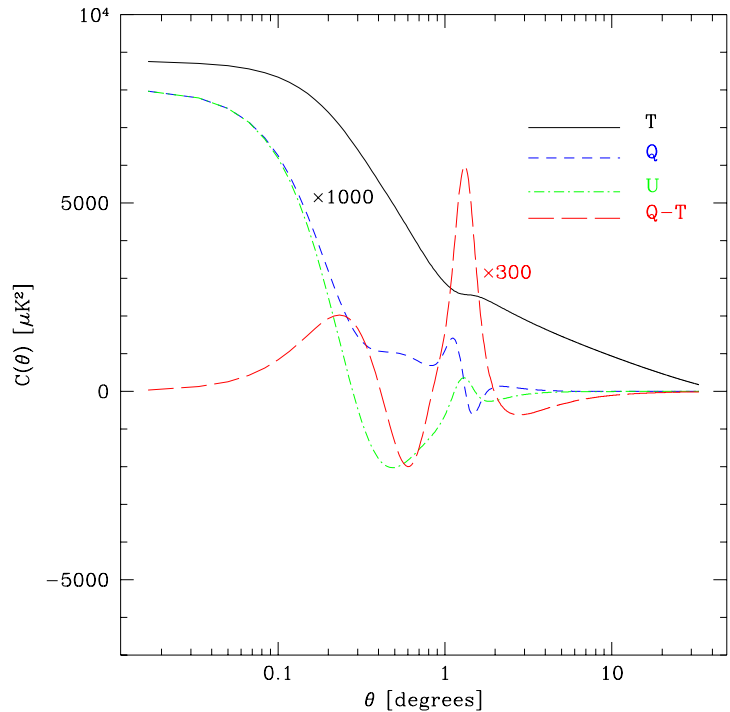}
\end{center}
\caption{Correlation functions in real space for COBE normalized standard
CDM model. The spectra have been smoothed with a $\theta_{fwhm}=0.2^o$
corresponding to the beam size of MAP.}
\label{figcorrcdm}
\end{figure*}

Figure \ref{figcorrcdm} shows the correlation functions in real
space. The spectrum has been smoothed with a $\theta_{fwhm}=0.2^o$
gaussian, similar to the MAP beam. The correlation functions are
defined in their natural coordinate system.
In this frame there is no cross-correlation 
between $U_r$ and $T$.
An interesting point is that both
polarization auto-correlation functions are negative for some angular range, which does not happen for the temperature.
To interpret the cross correlation we can
consider the polarization pattern around a hot spot ($T>0$). The cross
correlation starts positive, implying a radial  pattern of
polarization. Not all the polarization is correlated with the
temperature so it is difficult to see this trend in 
plate 3. In plate 4
we only plot the correlated part of the
polarization and here one can see more clearly 
that polarization ``vectors'' are preferentially
radial around hot spots.

Moving to larger angular scales the cross correlation changes sign.
When it is
negative the pattern becomes tangential. For large separations the
polarization around a hot (cold) spot is tangential (radial).
A point worth noting is that
the cross correlation vanishes at $\theta=0$, in
contrast to what happens for the other correlation functions.
This follows simply from symmetry, since there is no preferred 
direction that polarization should take. Only when we consider two points separated
by some distance is  the symmetry broken. The vector joining the
two points becomes the privileged direction and the polarization can be
preferentially parallel or perpendicular to this direction.

\section{Conclusions}
Polarization has a rich geometrical structure, which can be 
simply understood using a real space construction of scalar and 
pseudoscalar fields. These can be constructed as integrals over
Stokes $Q$ and $U$ parameters. Finite extent versions exist
which allow one to search for $B$ polarization without measuring
the whole sky.

Generation of polarization requires both Thomson scattering and 
significant quadrupole moment of photon distribution in electron
rest frame. We give simple arguments why this is so and discuss
predictions for some realistic models. Realistic attempts 
to extract polarization information have to include complications 
such as instrument noise and foregrounds and are given elsewhere
in these proceedings \cite{bouchet}.

\end{document}